\documentclass{article}

\usepackage{hyperref}
\usepackage[utf8]{inputenc}
\usepackage[T1]{fontenc}
\usepackage{xspace}
\usepackage{color}
\usepackage{changepage}
\usepackage{titlesec}
\usepackage{authblk}
\usepackage{lineno}
\usepackage{threeparttable}
\usepackage{amsmath}
\usepackage{amsfonts}
\usepackage{bm}
\usepackage{bbm}
\usepackage{graphicx}
\usepackage{fancyhdr}
\usepackage[backend=biber,url=false,doi=false,isbn=false,date=year, maxcitenames=2, minbibnames=2]{biblatex}
\usepackage[paper=a4paper,margin=1in, top = 20mm]{geometry}
\usepackage{subfiles}
\usepackage{mwe}
\DeclareFieldFormat{annotation}{%
  \newline\textbf{*}~\textbf{#1}%
}

\renewbibmacro*{finentry}{%
  \iffieldundef{annotation}{}{\printfield{annotation}}%
  \finentry%
}

\author[1,2]{Nicole Kleman$^{*}$}
\author[3,4]{Meng Lin$^{*}$}
\author[3,4]{Christopher R. Gignoux$^{*}$}
\author[2,5]{Arslan A. Zaidi$^{*\dagger}$}

\affil[1]{Bioinformatics and Computational Biology Program, University of Minnesota}
\affil[2]{Department of Genetics, Cell Biology, and Development, University of Minnesota}
\affil[3]{Department of Biomedical Informatics, University of Colorado Anschutz}
\affil[4]{Colorado Center for Personalized Medicine, University of Colorado Anschutz}
\affil[5]{Institute of Health Informatics, University of Minnesota}

\affil[$^{*}$]{Contributed equally}

\affil[$^{\dagger}$]{Correspondence to AAZ: aazaidi@umn.edu}

\title{Tread lightly interpreting group differences in genetic risk}

\date{April 2026}

\begin{document}

\maketitle

\section*{Abstract}

Observed differences in mean phenotypic values across human groups have attracted renewed interest with the rise of large-scale genomic studies and polygenic risk prediction. However, the genetic basis of these differences is far more difficult to establish than is often appreciated. Populations can diverge in allele frequency differences without diverging in mean genetic value. Empirical approaches to infer whether populations differ in mean genetic value fall under two broad categories: top-down approaches, which quantify the proportion of phenotypic variance explained by ancestry and bottom-up approaches, which compare polygenic scores across groups. However, both approaches have limitations that prevent them from reliably distinguishing true differences in genetic  apart from statistical artifacts like population structure, ascertainment bias, and poor cross-ancestry portability. Further, observed phenotypic shifts between populations may reflect bias in phenotype measurement and heterogeneity in study design rather than underlying genetic drivers. We argue that claims about group differences in genetic risk should be interpreted with considerable caution.

\section*{Introduction}

Observed differences in mean phenotypic values across human groups -- often defined by race or ethnicity but also genetic similarity -- have long been a source of scientific interest and public debate \cite{Feldman1975, Risch2002, Schraiber2024}. However, while group-level differences in disease prevalence and complex phenotypes are well-documented, the extent to which they are driven by genetic variation between groups is poorly understood. In public health and medicine, attempts to attribute group differences to genetic variation often reflect a desire to understand the mechanisms of health disparities. In other, more controversial domains, such attempts are sometimes tied to political projects. Recent advances in statistical methodology and availability of large-scale genomic datasets have brought renewed empirical attention to these questions \cite{AOUS, ELGH}. Yet confusion remains — even among specialists — about what genetic differences between populations do and do not imply and what we can actually learn about group differences from genetic data. In this review, we attempt to clarify these subtleties and highlight the formidable obstacles to inferring the contribution of genetic variation to phenotypic differences between groups — confusions that are rarely addressed, even in technical treatments of the subject.

\section*{Genetic differences between populations do not imply differences in mean genetic value}

A common argument in discussions of group differences is that because complex traits are heritable and allele frequencies vary between populations, genetic influences on those traits must necessarily differ between populations as well \cite{Reich2018}. The implication being not only that populations differ in allele frequencies, but that they differ in mean genetic value -- the sum of additive genetic effects across causal loci \cite{Lynch1998}. However, while the premise of this argument is not entirely wrong, the conclusion does not follow. What is often overlooked is that, at least for polygenic traits, allele frequency differences need not produce a difference in mean genetic value. In fact, in the limit of very high polygenicity, i.e., the infinitesimal model, the genetic variance between populations is dominated not by allele frequency differences but by linkage disequilibrium (LD) \cite{, Bulmer1971TheVariability, Latta1998DifferentiationTraits, Huang2025}. 

\begin{figure}[htbp]
\centering
\includegraphics[width=1\linewidth]{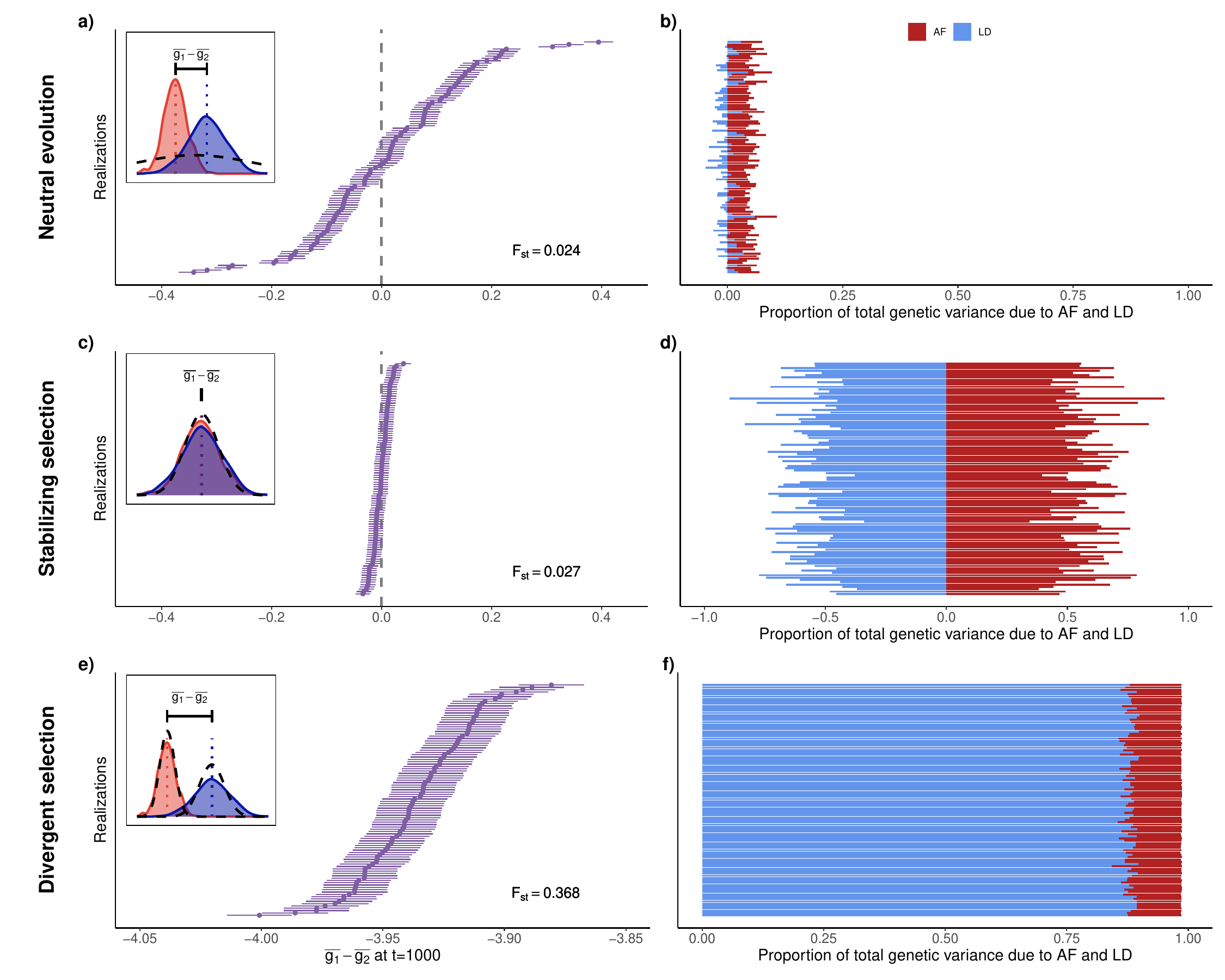}
\caption{\textbf{The expected genetic value for a trait under different evolutionary scenarios.} Simulations of two populations that split from an ancestral population ($t=0$) after 1,000 generations ($t$) for traits under neutral evolution (\textbf{a, b}), stabilizing selection (\textbf{c, d}) and divergent selection (\textbf{e, f}). (\textbf{a, c, e}) Populations' genetic values at generation 1,000, ordered on the y-axis by the magnitude of the difference of mean genetic values between two populations ($g_1$ and $g_2$) over 100 realizations with 95\% confidence intervals. Note that the x-axes in panels \textbf{a}, \textbf{c}, and \textbf{e} are on different scales reflecting the raw genetic value under each selection regime. Insets illustrate the optimal fitness in each population (black dashed line) and one realization of the distribution of genetic values. The mean $F_{st}$ across simulations is also listed on the bottom right. (\textbf{b, d, f}) At generation 1,000, the proportion of total genetic variance that is explained by allele frequency differences (red) and LD (blue) over 100 realizations (Supplement) \cite{Huang2025}. Note, that the LD contribution can be negative, especially under stabilizing selection \cite{Bulmer1971TheVariability}.}
\label{fig:exp_gen_value}
\end{figure}

To illustrate, we quantified the extent to which allele frequency differences between populations ($V_{\Delta f}$) contribute to the total genetic variance for a polygenic trait simulated under different selective constraints (Fig. \ref{fig:exp_gen_value}). These simulations are instructive in several ways. First, allele frequency differences contribute to genetic variance, even in the absence of any difference in mean genetic value. In fact, for a given trait, the difference in mean genetic value between populations is uncorrelated with $V_{\Delta f}$. Second, $V_{\Delta f}$ accounts for relatively little of the total genetic variance -- 4.8\% on average for neutrally evolving traits and 10.7\% for traits under directional selection (Fig. \ref{fig:exp_gen_value}b,f). Third, and perhaps counterintuitively, $V_{\Delta f}$ contributes a larger share of the total genetic variance under stabilizing selection than under directional selection or neutrality.

These results follow from the fact that the genetic variance between populations depends not only on the magnitude of allele frequency differences at causal loci, but also the direction of those differences with respect to each other. A population enriched for trait-increasing alleles at some loci might be depleted for such alleles at other loci, relative to another population. This \textit{directional LD} allows populations to diverge in allele frequencies without diverging in mean genetic value \cite{Lynch1998, Huang2025}. For polygenic traits under selection, directional LD contributes more to between-population genetic variance than allele frequency differences \textit{per se} (Fig. \ref{fig:exp_gen_value}e,f), which is precisely why tests of polygenic selection look for coordinated changes in allele frequency across loci rather than frequency differentiation at any given locus \cite{Pritchard2010, berg_population_2014, Mathieson2015}. Under stabilizing selection, the maintenance of similar trait means between populations requires that allele frequency differences be offset by negative LD across causal loci (Fig. \ref{fig:exp_gen_value}c,d). As a result, $V_{\Delta f}$ appears to account for a larger share of the total genetic variance, not because its absolute contribution is larger.

Taken together, these results underscore a simple, but underappreciated point: allele frequency differences are necessary but not sufficient to produce a difference in mean genetic value. The intuition that heritability plus differentiation implies a divergence in mean genetic value assumes that allele frequency differences are consistent in direction across loci -- an assumption that is violated for many genetic architectures, especially for neutral traits and traits under stabilizing selection. Establishing that populations differ in mean genetic value for a given trait therefore requires evidence beyond frequency differentiation at associated loci. In the sections that follow, we ask whether existing empirical approaches can provide that evidence.

\section*{Inferring differences in mean genetic value: promises and limitations}

The preceding sections raise the natural question: what evidence \textit{would} suffice to establish that two populations differ in mean genetic value for a given trait? This question is difficult to answer in naturally occurring populations, where differences in genetic value are confounded with environmental differences. Historically, quantitative geneticists circumvented this problem through common-garden experiments, in which organisms of different strains are raised under identical or randomized environmental conditions \cite{vanwallendael_common_garden_2022}. In such experiments, any systematic differences in mean phenotypic value between strains can be attributed to a difference in mean genetic value. In humans, where such experiments are neither practical nor ethical, researchers often rely on observational approaches that fall broadly into two categories depending on the direction of inference. \textit{Top-down} approaches start from the phenotype and attempt to attribute some portion of the variance between populations to genetic effects. \textit{Bottom-up} approaches work in the opposite direction by starting from trait-associated variants and their effects learned through genome-wide association studies (GWAS) and predicting and comparing mean genetic values across groups \cite{Abdellaoui2023}. As we discuss below, these approaches face distinct and complementary limitations.

\subsection*{Top-down approaches}

Top-down approaches infer differences in mean genetic value from patterns of covariance between phenotype and genetic ancestry, without identifying the causal variants. Two such approaches are commonly used: ancestry-trait correlations, which test whether genome-wide ancestry correlates with phenotypic value as evidence that differentiated alleles contribute to the trait mean, and local ancestry heritability methods, which estimate the total contribution of ancestry-differentiated loci to phenotypic variance \cite{Chan2023}.

Ancestry-trait correlations for polygenic traits are difficult to interpret for two main reasons. First, genetic ancestry is not randomly distributed with respect to the environment. Social determinants of health such as systemic racism, socioeconomic status, and access to healthcare all correlate with ancestry in admixed groups, driving health inequality \cite{martin_clinical_2019}. One study of health disparities found that Black Americans lost more years of their lives compared with White Americans for 28/36 physiological causes of death, even after adjusting for sex, age, and years of education \cite{Wong2002, Edge2016, Rosenberg2019}. Even within admixed African Americans, individuals with more genetic similarity to African populations were more likely to be diagnosed with 28/30 diseases that were significantly correlated with genetic ancestry (binomial p-value = $2.9 \times 10^{-8}$) \cite{Zaidi2023}. This consistent pattern of worse health outcomes is difficult to explain under a model where most traits are neutral or under stabilizing selection, especially since there is limited empirical evidence that some human populations have a higher burden of deleterious alleles than others \cite{Simons2014, Simons2016, Stolyarova2025TheGroups}.

A more subtle but fundamental problem is that ancestry-trait correlations are largely uninformative about whether the underlying signal can be attributed to (large-effect, highly differentiated) causal variants. With the exception of some traits such as prostate cancer, end-stage kidney disease, neutrophil count, and skin pigmentation -- whose architectures are dominated by one or a few large-effect, highly differentiated variants -- ancestry-trait correlations have not led to the discovery of variants with sufficiently large effects to explain the observed correlations \cite{Genovese2010, nalls_2008, Beleza2013}. This pattern is not surprising. As discussed above, group differences in trait mean are unlikely to be driven by large allele frequency differences at individual loci, but by the directional covariance in allele frequency differences across loci, which manifests as long-range directional LD in admixed populations. Thus, even setting aside environmental confounding, the genetic component of ancestry-trait correlations is expected to be dominated by long-range directional LD – itself a function of admixture history, including time since admixture, extent of gene flow, and assortative mating \cite{Zaitlen2017, Kim2021} -- and largely uninformative about allele frequency differences at causal loci.

One might consider estimating the genetic variance due to local ancestry \cite{Zaitlen2014}, which is less confounded with the environment than global ancestry, as a way of measuring the contribution of genetic ancestry to trait differences \cite{Shriner2015, Zaidi2017}. However, we showed recently that local ancestry heritability estimates measure the component of genetic variance arising from allele frequency differences between the source populations ($V_{\Delta f}$) but not the component arising from directional LD \cite{Huang2025}. As discussed earlier, $V_{\Delta f}$ typically explains a small share of the total genetic variance for polygenic traits and can be uninformative about differences in mean genetic value. In fact, local ancestry heritability can be non-zero in admixed populations, even in the absence of a difference in mean genetic value between the source populations \cite{Huang2025}. Thus, local ancestry heritability should not be interpreted as the heritability along the ancestry axis. Taken together, top-down approaches are fundamentally limited in their ability to say whether there is a difference in mean genetic value between populations \cite{Schraiber2024}.

\begin{figure}
    \centering
    \includegraphics[width=0.75\linewidth]{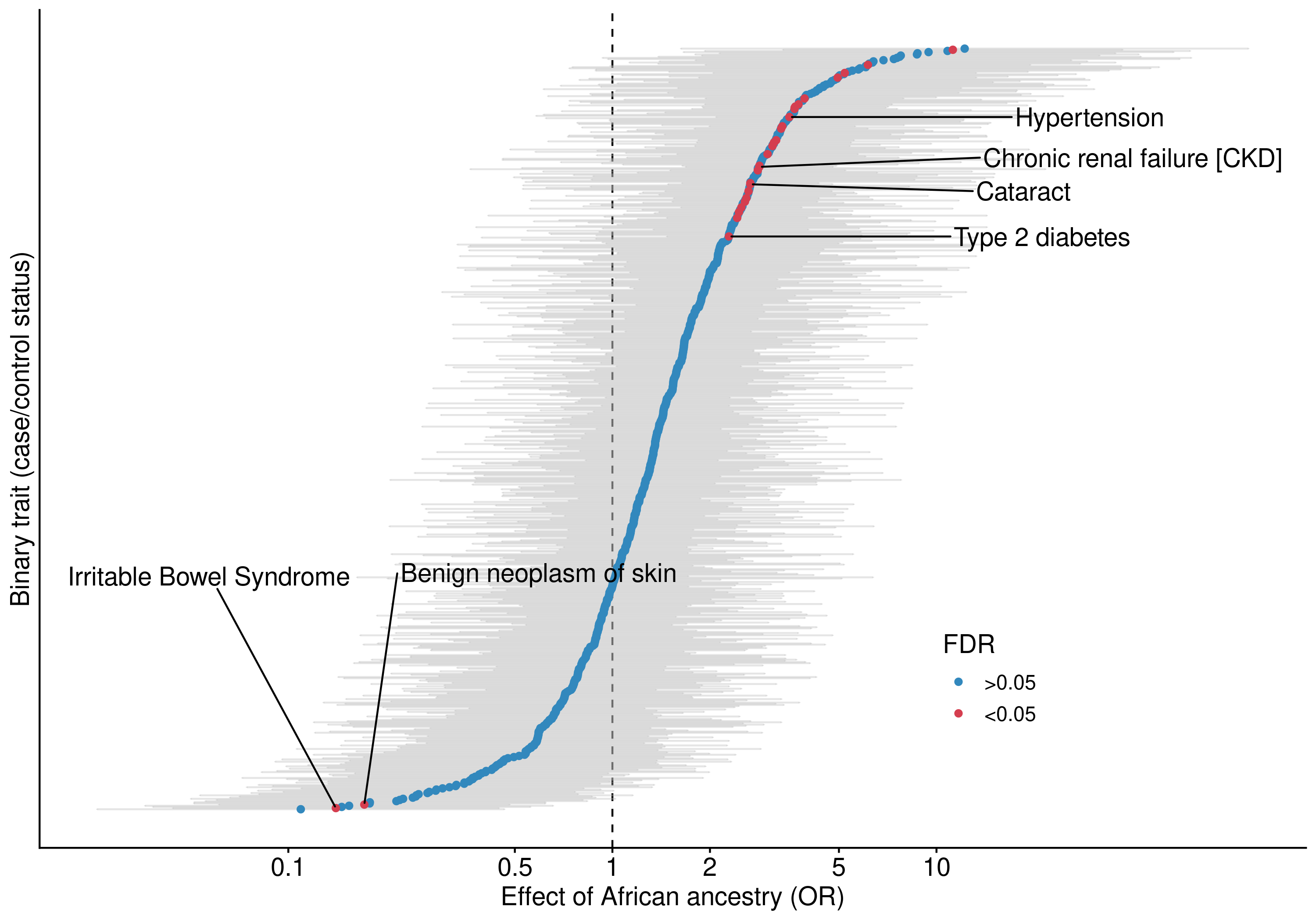}
    \caption{\textbf{Relationship between genetic similarity to Africans and 657 binary health-related traits among admixed African Americans.} The data are from \cite{Zaidi2023} generated under the following model: $logit(\pi) = \beta_0  + \beta_1\text{sex} + \beta_2\text{age} + \beta_3\text{age}^2 + \beta_4\text{nuclear ancestry} + \beta_5\text{mtDNA haplogroup} + \beta_6\text{mtDNA haplogroup}\times\text{nuclear ancestry}$. The traits (y-axis) are ordered by the effect of nuclear ancestry (x-axis) represented as odds ratio on the log-scale.}
    \label{fig:pmbb_ancestry}
\end{figure}

\subsection*{Bottom-up approaches}

In principle, if we knew the causal variants and had unbiased estimates of their effects, we could predict the genetic values of all individuals and directly estimate the difference in mean genetic values between populations. This is precisely what polygenic scores (PGS) attempt to do: aggregate the estimated effects of trait-associated variants to predict the genetic value for each individual. In practice, however, standard PGS approaches fall short of this ideal in several important ways, exhibiting poor generalization of trait prediction across ancestries. This leads to challenges in interpreting between-group difference in PGS \cite{coop2019reading} and the harm versus benefit in deploying a one-size-fits-all ancestry calibration to compare PGS across ancestries. Interpreting differences between-groups in PGS is complicated by two related problems. First, the prediction accuracy degrades with the genetic distance from the training population \cite{martin_clinical_2019, ding_polygenic_2023}. Second, mean PGS distributions shifts between groups in ways that may not reflect true differences in mean genetic value \cite{martin_human_2017}.

These problems arise from a set of known causes. First, GWAS, upon which a PGS model is built, have not recovered the full genetic architecture of complex traits -- the `missing heritability' problem \cite{manolio_finding_2009}. Thus, between-group differences in PGS represent information from a subset of the loci and there is no guarantee that the remaining, undiscovered portion of the architecture will yield the same pattern. Second, genotyping arrays and imputation panels tag causal loci imperfectly and unequally across ancestries, leading to ancestry-related differences in variant discovery and effect size estimation \cite{novembre_tread_2018, ding_polygenic_2023}. The choice of LD panel for PGS training (either for thinning markers or shrinking effect sizes) compounds this problem. Third, residual population stratification in GWAS inflates between-ancestry differences in PGS distributions, which might capture environmental differences rather than true differences in mean genetic value \cite{berg_reduced_2019, Sohail2019a, zaidi_demographic_2020, Blanc2025}. Beyond these technical limitations, standard PGS approaches assume additive effects and may fail to capture non-linear relationships, gene-by-environment interactions, and epistasis \cite{mathieson_omnigenic_2021, coop2019reading}. How these limitations affect PGS portability across ancestries is largely unknown and an area of active research.

\begin{figure}[htbp]
\centering
\includegraphics[width=1\linewidth]{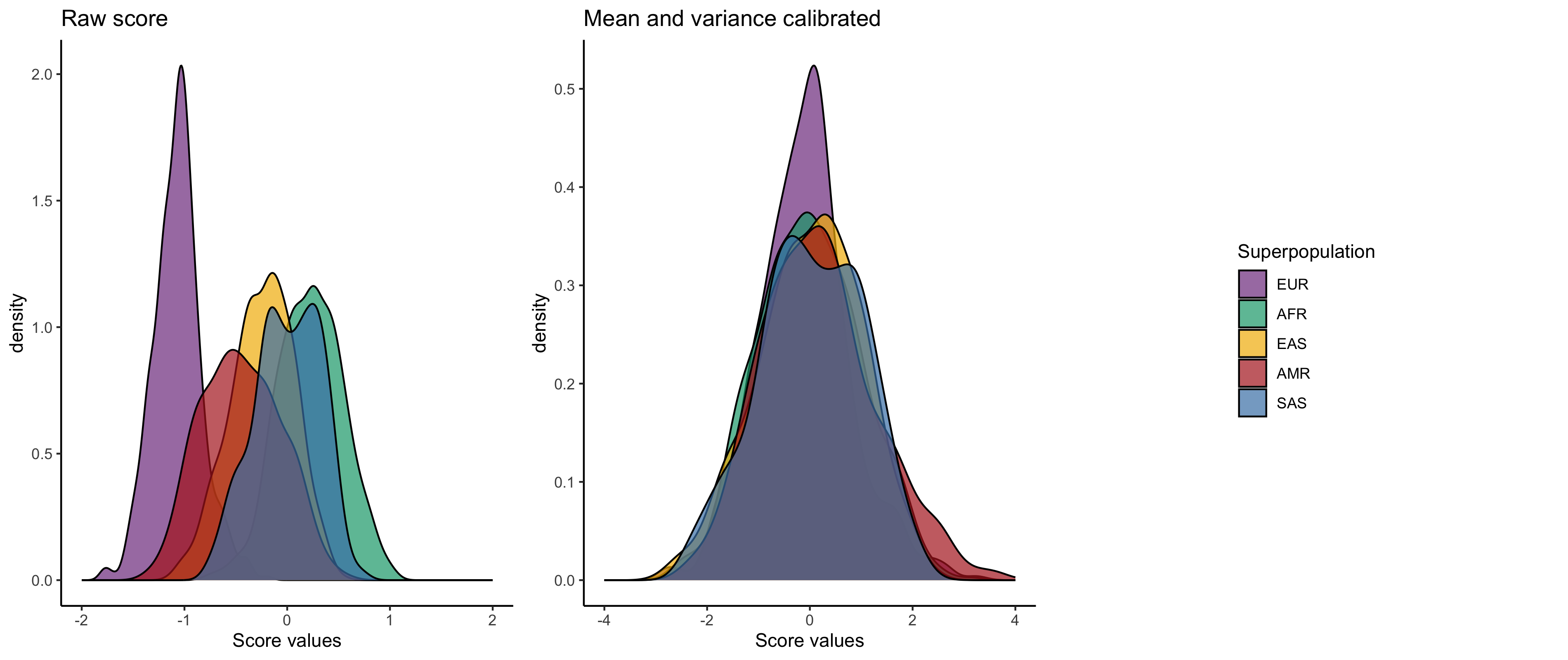}
\caption{\textbf{Distribution of a skin pigmentation score across continental groups prior to and after ancestry calibration} (PGS002110 \cite{prive_portability_2022,lambert_polygenic_2021}). Samples are from 1000 Genomes Project (high coverage \cite{byrska-bishop_high-coverage_2022}). Scores are shown in uncalibrated raw values (left) and after ancestry calibration of both mean and variance using first 5 PCs (right).}
\label{Figure 2}
\end{figure}

Developments are underway to rectify the ancestry bias problems. First, better representation and modern platforms such as Global Diversity Array and whole-genome sequencing will improve tagging of causal variants to reduce ancestry-specific ascertainment in GWAS \cite{wojcik2018imputation,martin2021low}. Second, \textit{post hoc} ancestry calibration methods enable PGS comparisons between groups by adjusting for biases that track with ancestry \cite{khan_genome-wide_2022,ge_development_2022, lennon_selection_2024, lambert_enhancing_2024, Hou2024}. The second approach of ancestry calibration models the mean and/or variance of a PGS distribution as a linear function of ancestry or principal components (PCs), making specific assumptions of how ancestry impacts PGS. Despite rising popularity, it is unclear whether these assumptions are justified or whether the adjustments introduce additional distortions. In fact, for traits with a true shift in mean genetic values between populations, the standard calibration approach may make PGS distributions appear overly similar. These presumably are traits that underwent a combination of directional selection, local adaptation, recent migration bottlenecks, or other forces resulting in large $F_{ST}$ at causal loci. A well-characterized example of skin pigmentation illustrates this clearly: despite being highly heritable and strongly differentiated across populations \cite{Beleza2013, Lamason2005, Crawford2017, Martin2017}, standard calibration procedures remove all group differences in PGS (Fig. \ref{Figure 2}). Researchers would benefit from understanding the sources and structure of bias in their specific setting before attempting to use calibration methods.

In conclusion, current polygenic scores can still be imprecise and biased, and the direction of bias varies across populations in unpredictable ways that may not be `rescued' by current methods in a straight-forward way. Skin pigmentation remains a conspicuous example, yet how many other traits affected by similar processes but less easily detected\cite{harpak2021evolution}? Until the sources of bias in polygenic scores can be adequately characterized and corrected for, bottom-up approaches, despite their promise, cannot reliably distinguish true differences in mean genetic value from artifacts of discovery, portability, and stratification.

\section*{Phenotype measurement introduces additional barriers to inference}

Beyond the limitations described above, a more fundamental problem is that the phenotypes being analyzed represent a marginal aggregate over genetic and environmental processes. We measure phenotypic variation in the context of the individuals, populations, environment, and study design, impacting genetic inference and prediction even within populations \cite{mostafavi2020variable}, let alone across populations or in diverse, multi-ancestry cohorts, which are highly structured and complex. Therefore, the problem of identifying differences based on genetics alone cannot be simply addressed by improving our understanding of genetic variability across populations in an epidemiological vacuum. Beyond population structure, trait architecture inference is driven by differences in study design (covered elsewhere, e.g., \cite{CUDIC2026}), ascertainment \cite{Kim2018, vanAlten2025}, and potentially more importantly, differences in phenotypic measurement driven by patterns in healthcare usage or clinical recommendations. 

There are multiple clinical measures that, whether known or unknown to the researcher, imbue systematic differences that may not result from genetics at all. Revisiting these measures, particularly in clinical settings, has recently become an active area of re-consideration by professional societies in commonly-measured quantitative traits such as pulmonary function \cite{bhakta_race_2023} and expected glomerular filtration rate (eGFR) \cite{delgado_unifying_2021}. This is also true for wide-ranging disease domains including cancer \cite{esnaola_racial_2012}, clinical actionability of self-reported pain \cite{hoffman_racial_2016}, and osteoporosis \cite{cauley2011defining}. The concern has been that race-based clinical reference ranges could reflect differences in unmeasured environmental exposures rather than underlying genetic differences. Even standard-of-care specific clinical variables could recapitulate these biases, reinforcing historical race-based essentialism (e.g., \cite{braun_race_2023}). However, this is not unidirectional and other groups have demonstrated the utility of this approach and how it can be performed to alleviate, rather than exacerbate, known health disparities \cite{borrell_2021}. Thus, how a phenotype is measured, and when, becomes an equally critical part of studying health disparities as statistical methods. 

Measurement bias has consequences that extend beyond genetic inference. Biased measurements result in fewer clinical referrals \cite{landon_assessment_2021}, delayed diagnosis, are more severe disease on initial diagnosis \cite{langer-gould_racial_2025}. As the field of large-scale genetic epidemiology has moved from standardized phenotype measurement to studies of convenience based on electronic healthcare records and surveys \cite{hu_variations_2023}, measurement bias becomes harder to detect and account for. Awareness of how study design affects these biases is essential to work towards fair phenotypic measurement across groups. 

\section*{Conclusion}

Where does this leave us? There is much promise in studying genetic architecture in larger and more diverse cohorts, and our discussion should not discourage this line of research. On the contrary: structured and admixed populations offer unique opportunities to ask questions that may not be possible in homogeneous settings. As genetics becomes increasingly integrated into clinical and public‑health settings, understanding how genotype maps to phenotype across human diversity will only grow in importance. 

However, promise does not eliminate responsibility. Interpreting group differences requires particular care, because current methods are limited in ways that can blur the line between true biological signal and statistical artifacts of clinical record, study design or population structure. In this context, precision in language is not merely academic: careless phrasing can be misleading and cause social harm by reinforcing notions of genetic essentialism. 

Moving forward, progress will depend not only on better data and better models, but also on deliberate, precise communication about what our tools can -- and cannot -- reveal about human variation. Statistical genetics is a rapidly evolving field and our inference of genetic architecture is provisional, contingent on current datasets, methods, and assumptions. Being explicit about this uncertainty -- teaching it, modeling it, and communicating it -- is essential for scientific integrity and public trust.

\section*{Code Availability}
We carried out all analyses in R version 4.5.1 or 4.2.2 \cite{R2023} and SLiM 5.1 \cite{SLiM}. PGS calculation was done using ESCALATOR \url{https://github.com/menglin44/ESCALATOR}. All code is freely available on \url{https://github.com/zaidilab/COGEDE}.

\section*{Acknowledgements}

This work was supported by National Institutes of General Medical Science award R35GM159945 and R00GM137076 to AAZ and National Human Genome Research Institute award U01HG011715 to CG. The content of this paper is solely the responsibility of the authors and does not necessarily represent the official views of the NIH. We thank Doc Edge for providing helpful feedback.

\newpage

\printbibliography
\newpage

\renewcommand\thefigure{S\arabic{figure}}    
\setcounter{figure}{0}  

\section*{Supplement}

\subsection*{Methods}

\textbf{Simulations for expected genetic values:} Using SLiM \cite{SLiM}, we simulate a polygenic trait forward in time under neutral evolution, stabilizing selection, and divergent selection. First, we initialized a (burn-in) randomly mating population of 10,000 diploids for 40,000 generations with 10 chromosomes, each of 100kb, mutation rate of $1 \times 10^{-07}$, and recombination rate of $1 \times 10^{-08}$. Next, we split this `ancestral' population into two populations of equal size (10,000) and let them evolve for the next 1,000 generations. The effect of new mutations on the trait were drawn from $\mathcal{N}(\mu_{\beta} = 0, \sigma_{\beta}^2 = 0.005)$, following \cite{Schraiber2024}. We model selection on the individual level with Gaussian fitness functions. Briefly, the relative fitness of an individual with trait value $x$ in an environment with an optimum value (fitness peak) of $\mu_o$ and variance of $\omega_o$ is $W(x | \mu_o, \omega_o ) = exp(-\frac{x - \mu_o}{2\omega_o})$. For a trait under stabilizing selection, the fitness function becomes $\mathcal{N}(\mu_o = 0, \omega_o = 16)$ before and after the population split. To simulate a trait under divergent selection the fitness function is the same as stabilizing selection during the 40,000 generation burn-in, and changes to $\mathcal{N}(\mu_o = -2.0, \omega_o = 16)$ and $\mathcal{N}(\mu_o = 2.0, \omega_o = 16)$ for populations 1 and 2, respectively, after the split. For a neutral trait, the fitness is uniform, i.e., independent of trait value. Each evolutionary scenario was replicated 100 times. To calculate the proportion of genetic variance due to allele frequency differences, we computed $\frac{\sum^m_{i=1} \beta^2_i (f^1_i - f^2_i)^2}{V_g}$; where $V_g$ is the total genetic variance in the two-population system, $m$ is the total number of causal loci, $\beta_i$ is the effect size of the $i$th locus, $f^1_i$ and $f^2_i$ are the allele frequencies in populations 1 and 2, respectively \cite{Huang2025}. $V_g$ was calculated as the variance of the genetic vlaues, $g$. To compute the proportion of genetic variance due to LD, we first calculated the genetic variance between populations as $V_{gb} = \frac{(\bar{g_1} - \bar{g_2})^2}{4}$ and then subtracted out the variance due to allele frequency: $\frac{V_{gb} - \sum^m_{i=1} \beta^2_i (f^1_i - f^2_i)^2}{V_g}$; where $\bar{g_1}$ and $\bar{g_2}$ are the mean genetic values for populations 1 and 2, respectively.

\noindent \textbf{Correlations between genetic ancestry and binary traits:} To illustrate the pattern of correlation between health-related traits and genetic ancestry among admixed Americans, we downloaded Table S4 from \cite{Zaidi2023}. This table reports the association between proportion of African ancestry and 1,191 traits among 8,311 individuals with mixed African and European ancestry from the Penn Medicine Biobank. To fit this association, the authors used logistic regression for binary traits (phecodes) and linear regression for quantitative traits with nuclear ancestry, mtDNA ancestry, the interaction between the two, and sex, age, and age\textsuperscript{2} as the independent variables. From this set, we retained the results of 657 binary traits which had at least 100 cases.

\noindent \textbf{PGS calculation in 1000 Genomes Project:} We obtained high coverage 1000 Genomes Project (TGP) in VCF format \cite{byrska-bishop_high-coverage_2022} from \url{https://ftp.1000genomes.ebi.ac.uk/vol1/ftp/data\_collections/1000G_2504_high_coverage/working/20220422_3202_phased_SNV_INDEL_SV/}, and converted the data to PLINK2 format \url{(www.cog-genomics.org/plink/2.0/)} \cite{chang2015second}. We restricted the samples to 2,504 unrelated individuals commonly present in earlier releases of TGP. A skin pigmentation score weight file (PGS002110 \cite{prive_portability_2022}) was downloaded from PGS Catalog \url{(https://www.pgscatalog.org)} \cite{lambert_polygenic_2021}, and applied to all individuals for score calculation through ESCALATOR \url{(https://github.com/menglin44/ESCALATOR)}. 275,451 out of 275,831 variants from the original model ended up in score calculation, after the harmonization step excluding 6 variants with mismatched allele codes and 374 variants missing in TGP.

\noindent \-textbf{Ancestry calibration of PGS:} Principal component analysis was performed using PLINK2 --pca command \cite{chang2015second, Purcell2007}, after pruning out variants with missing rate >5\%, minor allele frequency <1\%, LD >0.1 (--indep-pairwise 500 125 0.1), and aggregating markers across autosomes. We calibrated the original PGS values by modeling PGS mean and variance as a linear function of the top 5 PCs following methods by \cite{khan_genome-wide_2022, ge_development_2022} (see code availability). Specifically, PRS values can be modeled as $PGS_{obs} = \alpha_0 + \sum \alpha_iPC_i + \epsilon$, where the residual $\epsilon$ is equivalent to mean-calibrated PGS; further, variance is modeled as $\epsilon^2 = \beta_0 + \sum{\beta_iPC_i} + \epsilon'$, with mean and variance calibrated score as $\frac{PGS_{obs} - (\alpha_0 + \sum{\alpha_iPC_i})}{\sqrt{\beta_0 + \sum{\beta_iPC_i}}}$.

\subsection*{Supplemental Figures}

\begin{figure}[htbp]
\centering
\includegraphics[width=1\linewidth]{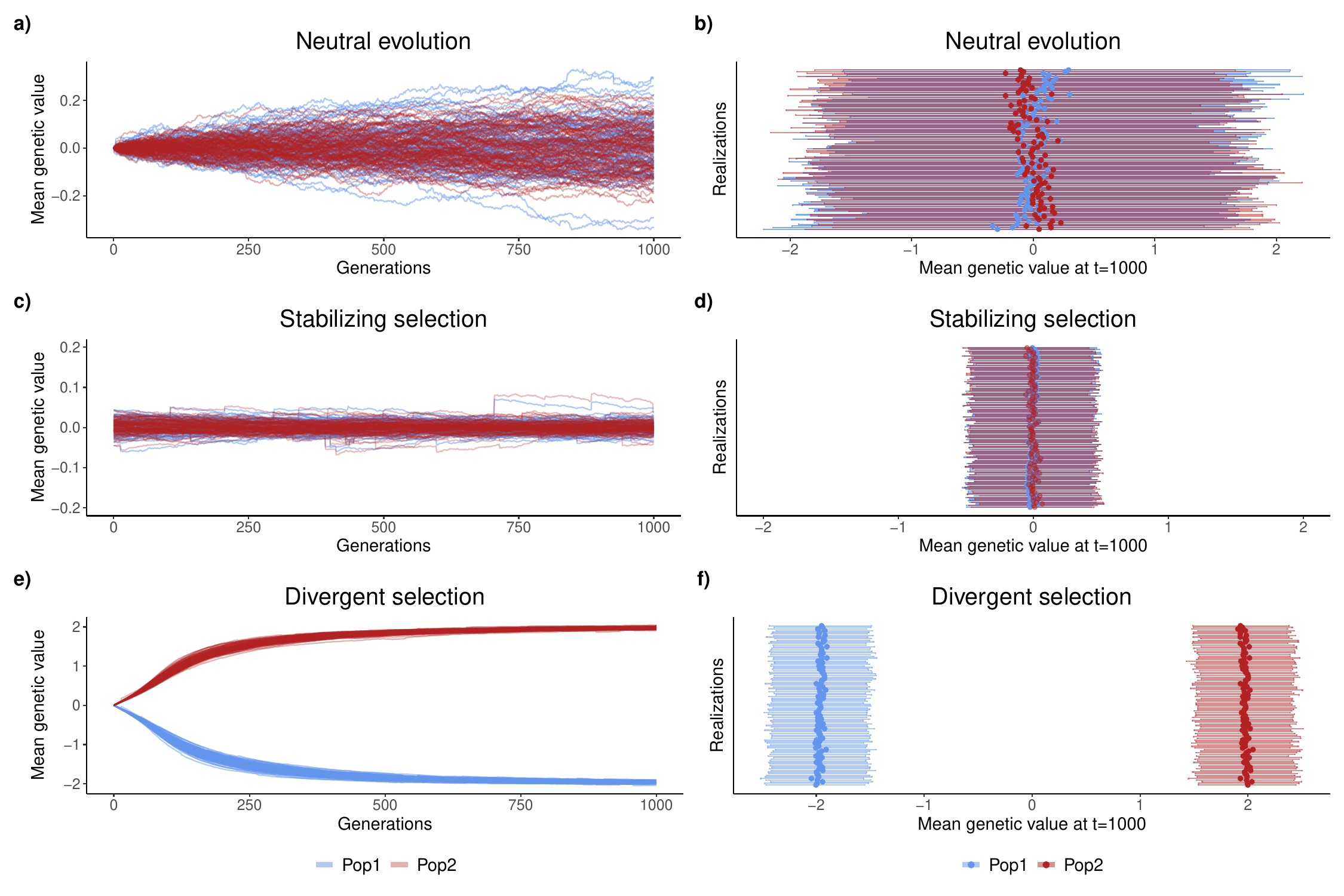}
\caption{\textbf{Mean genetic values of two populations across 1,000 generations for traits under different evolutionary scenarios.} (\textbf{a, c, e}) An ancestral population of 10,000 individuals splits into two populations of equal size (10,000) after 40,000 generations, then the selection type continues to act of the polygenic trait for an additional 1,000 generations. (\textbf{b, d, f}) At generation 1,000, the mean genetic value for two populations (population 1 in blue and populations 2 in red) ordered by the difference in magnitude between the two populations genetic values across 100 realizations with 95\% confidence intervals.}
\label{gen_var_supp}
\end{figure}

\end{document}